\documentclass[12pt,preprint]{emulateapj}

\newcommand{\beq}{\begin{equation}}
\newcommand{\beqa}{\begin{eqnarray}}
\newcommand{\eeq}{\end{equation}}
\newcommand{\eeqa}{\end{eqnarray}}

\begin{document}

\title{Galaxy Bulge Formation: Interplay with Dark Matter Halo and Central Supermassive Black Hole}

\author{Bing-Xiao Xu$^1$, Xue-Bing Wu$^1$ and HongSheng Zhao$^2$}

\affil{$^1$Department of Astronomy, Peking University, 100871
Beijing, China} \affil{$^2$SUPA, University of St Andrews, KY16 9SS,
Fife, UK} \email{xubx@bac.pku.edu.cn, wuxb@bac.pku.edu.cn,
hz4@st-andrews.ac.uk}

\begin{abstract}
We present a simple scenario where the formation of galactic bulges
was regulated by the dark halo gravity and regulated the growth of
the central supermassive black hole. Assuming the angular momentum
is low, we suggest that bulges form in a runaway collapse due to the
"gravothermal" instability once the central gas density or pressure
exceeds certain threshold (Xu \& Zhao 2007). We emphasize that the
threshold is nearly universal, set by the background NFW dark matter
gravity $g_{DM} \sim 1.2 \times 10^{-8}{\rm cm}\ {\rm sec}^{-2}$ in
the central cusps of halos. Unlike known thresholds for gradual
formation of galaxy disks, we show that the universal
"halo-regulated" star formation threshold for spheroids matches the
very high star formation rate and star formation efficiency shown in
high-redshift observations of central starburst regions. The
starburst feedback also builds up a pressure shortly after the
collapse. This large pressure could both act outward to halt further
infall of gas from larger scale, and act inward to counter the
Compton-thick wind launched from the central black hole in an
Eddington accretion. Assuming the feedback balancing inward and
outward forces, our scenario naturally gives rise to the black
hole-bulge relationships observed in the local universe.
\end{abstract}

\keywords{black hole physics -- galaxies: formation -- galaxies:
nuclei -- galaxies: starburst -- galaxies: structure}

\section{Introduction}

It is now widely believed that the violent star formation triggered
by the merger of gas-rich galaxies is related to the formation of
the galactic spheroidal component in the early Universe. The
observational counterpart of these formation events may be
represented by some intense star formation activity such as the
Lyman break galaxies (LBGs) and submillimetre galaxies (SMGs) at
$z\ge3$ \citep{ST96,SML97}, where the star formation rate could be
as high as $\sim1000M_{\odot}yr^{-1}$. These observed star-forming
galaxies are just the progenitors of the spheroid-dominated galaxies
in the local Universe. Their extremely high star formation rate
implies that a large amount of gas turns into stars in a relatively
short timescale, and the star formation efficiency (SFE) associated
with these extreme situations are also very high ($30-100\%$), much
higher than the SFE inferred by the Schmidt-Kennicutt star formation
law in the nearby disk galaxies and starburst galaxies \citep{KN98}.
It has been proposed that there may be two different star formation
modes for the disk and the spheroidal component respectively
\citep{SL05}. In the disk mode, the gravitational instability
\citep{KN89} and the porosity \citep{SL97} regulated mechanisms can
keep the disk star formation rate and SFE from raising too high; in
the spheroid mode, the star formation may be triggered by the bar
instability in relatively less massive galaxies
\citep{CM90,HPN93,WS04}; however, there is a lack of detailed models
for the star formation in the massive spheroid component where a
self-regulated mechanism is probably absent. So it has been
suggested that the star formation may proceed in a maximum way
during the formation of galactic bulges\citep{EG99}.

Many evidences also show that the growth of supermassive
black hole (SMBH) in galactic center is closely related to the star
formation activity \citep{AX03,HE05,PA04}. The similarity in the
comoving space density between the star-forming galaxies and quasars
offers strong proof for such a causal connection
\citep{CB03,PA04,SV04}. The tight relationship of the black hole
(BH) mass with the bulge velocity dispersion \citep{FM00,GB00,TMA02}
and the bulge mass \citep{MA98,MD02,MH03} also reveals such apparent
association. On the other hand, observations of some starburst
galaxies and bright quasars have shown clear evidences of outflows
\citep{PT00,AD03,PO03}. This indicates that the mechanical feedback
may play an important role in galaxy evolution and BH growth. Many
previous discussions have emphasized that the central BH can
interact with the surrounding environment in a self-regulated way
\citep{SR98,HNR98,BL99,FA99,KI03,WL03,MQT05,BN05}. All these models
are based on an "outward" nuclear feedback scenario and are
essentially similar; they differ only in whether the energy
deposition or the momentum deposition dominates. All these models
lead remarkably to the BH-bulge relation. Nevertheless,
few of these studies carefully considered the star formation
activity during the formation of the spheroid component and its
relation with the BH growth even though we know it is important.
Therefore, it seems still worth to revisit more details of this
topic.

The motivation of this paper is to link several physical
phenomena mentioned above, and explain them coherently.
In this paper, we will present a simple analytical model to describe
the star formation during the formation of the spheroidal component
and the possible connection between the central starburst and the
BH's growth. We show that the isothermal gas sphere embedded in a
NFW dark matter halo becomes unstable at some critical radius once
the ratio of the central gas pressure to the dark matter pressure
exceeds certain threshold value, and the gas within this critical
radius (maximum stable gas mass) will collapse and trigger the
central starburst. Intense star formation accompanies strong
star-forming feedback which plays an important role in resisting the
gravity at large scale; while at small scale, the inward star
forming feedback is introduced to connect the starburst region and
the nuclear region by virtue of the theoretical and observational
considerations. We also show that the BH's growth is confined and
regulated by the inward star-forming feedback. Based on such inward
star forming feedback scenario, the BH-bulge relations are naturally
derived, which well match with the observations. We also show
that the narrow range of the starburst duration and the dust-torus
structure in AGNs can be explained under the same scenario.

The paper is organized as follows. In \S2 we first discuss the
density profile of the background dark matter potential and the
isothermal gas sphere, then discuss the stability criterion of the
isothermal gas sphere embedded in the dark matter background in
detail. In \S3 we describe the effects of the star-forming feedback
and make a detail description of the so-called "halo-regulated" star
formation mode. We discuss the effect of the inward star forming
feedback and the obscured BH's growth in \S4, and eventually obtain
the BH-bulge relations in this section. We present some discussions
in \S5 and the main conclusions in \S6.

\section{GRAVOTHERMAL INSTABILITY AND PROTOGALAXY COLLAPSE}

Following \citep[hereafter XZ]{XZ07}, we model the protogalaxy as an
isothermal gas sphere embedded in the central $r^{-1}$ cusp region
of the dark matter potential, adopting the NFW density distribution
for the dark matter \citep{NFW97}. The dark matter density profile
is given by \beq \rho_{NFW}(r)\approx \Pi r^{-1}, \qquad \Pi \equiv
130 M_{\odot} {\rm pc}^{-2}M_{v,12}^{0.07}\zeta_c(z),\eeq where
$M_{v,12}=M_{vir}/10^{12}M_{\odot}$ and $M_{vir}$ is the virial mass
of the halo, $\zeta_c(z) \approx [0.3+0.7(1+z)^{-3}]^{2/3} \times
0.58 [\ln(1+c)-c/(1+c)]$ where
$c\approx13.4M_{v,12}^{-0.13}(1+z)^{-1}$ is the concentration
parameter \citep{BN98,BL01,BU01}. Such a dark matter distribution
produces almost a constant gravity \beq g_{\rm
DM}(r)=\frac{GM_{DM}(r)}{r^2} = 2 \pi G \Pi \sim 1.2 \times
10^{-8}{\rm cm}\ {\rm sec}^{-2}. \qquad \eeq

After the baryons entering the dark matter halo, they will be
shock-heated to the virial temperature. Then they will cool and
settle into the core region of the dark halo in gaseous form. If we
assume the angular momentum is low, an isothermal spherical
structure will be built in the core region of dark matter. Because
there is large amounts of gas accumulating in the central region,
the self-gravity of the gas sphere is not negligible. The combined
gravity from the gas sphere itself and the background dark matter
potential balances with the gas pressure. So we have the hydrostatic
equilibrium equation in the inner region \beq
{\sigma^2\over\rho}{d\rho\over dr}=-{G(M_{gas}+M_{DM})\over
r^2}=-{G\int (\rho+\Pi r^{-1})4\pi r^2dr\over r^2}, \eeq where
$\rho(r)$ is the gas density and $\sigma$ is the velocity
dispersion. The boundary conditions to be satisfied at the center
$r=0$ are \beq \rho(0)=\rho_0,\qquad {d\rho\over dr}=-{2\pi\Pi
G\rho_0\over\sigma^2},\eeq where $\rho_0$ is the central density.
Once $\rho_0$ and $\sigma$ are given, the gas density and mass
profile under the hydrostatic equilibrium can be totally determined.

The stability of gas sphere embedded in a background potential was
first investigated by \citet{SP42}, who pointed out that there is a
maximum stable gas mass for a given temperature with a background
potential of stars. A more recent sophisticated calculation was
provided by \citet{EG99}. He directly integrated the hydrostatic
equilibrium equations for a mixture of the cold gas and dark matter.
The equilibrium equations are $dP_{gas}/dr=\rho_{gas}g$ for gas and
$dP_{DM}/dr=\rho_{DM}g$ for dark matter. The gravitational
acceleration $g$ comes from the both components and can be
determined by the Poisson equation: $\nabla\cdot g=-4\pi
G(\rho_{gas}+\rho_{DM})$. He used various density profile for gas
and dark matter in his calculations, involving different central gas
density. He found that there is no solution satisfying the
 equations if the ratio of the central gas density to
the central dark matter density is larger than certain threshold
value. Above such threshold value, any additional mass added to the
combined components of gas and dark matter will make the whole
system unstable.  These two studies share common feature of finding
a maximum gas mass from the hydrostatic or virial equilibrium
equations. If the cumulated gas mass is larger than the maximum one,
the gas pressure can not support the combined gravity from the
sphere itself and the background dark matter, the sphere will
inevitably collapse. So such instability is mainly affected by the
"external" fluctuations. In another words, the equilibrious sphere
will always keep in equilibrium unless there is additional mass from
the outside added into the system.

The idea of XZ is similar but they emphasize the role of the so-call
"gravothermal" instability: consider an isothermal gas sphere with
fixed enclosed mass. A tiny compression of the sphere will decrease
the volume and increase the density everywhere. The isothermal gas
sphere would be stable with the density increasing everywhere at the
same time, because the pressure will increase everywhere to push
back the decreasing volume, balancing the increased gravity.
However, if the density increases at the center, but is reduced at
some large radii, the isothermal gas in the reduced pressure region
can not support the gravity, and the region is unstable. In another
words, the onset of instability is still possible even if there is
no additional mass added to the system. Actually, this is a
thermodynamic type of instability which originates from the negative
specific heat brought by the self-gravity of the sphere. The
gravothermal instability of the pressure-bounded isothermal sphere
has been investigated by many authors \citep{BN56,LW68,LB01}. It is
suggested that there is a threshold density contrast between the
center and the edge, above which the system will become unstable.
For the isothermal sphere embedded in a background potential, the
background potential provides an additional inward force which is
much like the external pressure in the pressure-bounded isothermal
sphere. However, it has not been seriously considered. Using the
variation method to investigate the problem, XZ have derived the
exact stability criterion for the isothermal sphere with a
background NFW potential. The criterion for the gas sphere to keep
stable is \beq p_0 \equiv \rho_0 \sigma^2 \le 31 p_{DM,0} \eeq where
$p_0$ is the central gas pressure and \beq p_{DM,0}=4\pi G\Pi^2 \eeq
is the central dark matter pressure. XZ find that the instability
starts at a critical radius $r_c\approx \sigma^2/ 2.2\pi\Pi G$,
inside which the enclosed gas mass is \beq
M_{g,max}=1.78\times10^{11}\sigma_{200}^4M_{v,12}^{-0.07}\zeta_c(z)^{-1}
M_{\odot}.\eeq XZ suggest that there is a threshold of the gas mass
$M_{g,max}$ to collapse and trigger rapid star formation. It is
reasonable to consider that the bulge formation just originates from
such a gas collapse  when the stability criterion Eq. (5) is
violated. The gas that collapsed due to self-gravity is the direct
material source for the bulge formation.

\section{THE OUTWARD STAR FORMING FEEDBACK DURING THE BULGE FORMATION}

In this section we will discuss the effects of the star-forming
feedback. First, we will give a brief description to the
star-forming feedback in starburst galaxies; Then we describe the
so-called "halo-regulated" star formation mode by considering the
effects of star-forming feedback. We note that many conclusions
inferred from such star formation mode are consistent with recent
observations. We point out that such mode is not only expected
theoretically, but also actually exists in some extreme starburst
regions.

\subsection{The effects of star-forming feedback}

When the central gas density exceeds the critical value, the
collapse will compress the gas sphere and enhance the gas density.
The squeezed gas is capable to form molecular clouds and eventually
form stars inside the clouds, and a large amount of the gas
accumulating in the central region will trigger the central
starburst. Vigorous star formation activities will inevitably lead
to strong star forming feedback. Here we mainly focus on two primary
sources of star forming feedback: radiation pressure and supernovae.
We know that the UV radiation can be absorbed by the dust
efficiently, so the radiation pressure of the newly formed stars becomes important if the optical depth to the dust $\tau_{Dust}$
can reach the unity. Actually, observations have showen that there
are a large amount of dusts in the local and distant starburst
galaxies \citep{LH96,SM96,AS00}. They should be produced by the
supernovae rather than the AGB stars because of the relatively short
duration of the starbursts. For a large star formation rate, the
timescale for supernovae to reach $\tau_{Dust}\approx1$ can be very
short in comparing with the starburst duration \citep{MQT05}. The
momentum deposition rate of radiation pressure can be expressed as
$\dot P_{rp}=\epsilon\dot M_{\star}c$, where $\dot M_{\star}$ is the
star formation rate,$\epsilon$ is the efficiency of converting mass
to energy in starbursts, and $\epsilon\approx10^{-3}$ for a Salpeter
IMF according to some starburst models \citep{LE99}. Here we assume
that the star formation rate is proportional to the gas mass
$M_{c,max}$ and inversely proportional to the dynamical time scale
$t_{dyn}$, \beq \dot M_{\star}=\eta\frac{M_{g,max}}{t_{dyn}} ,\eeq
where $\eta$ is the star formation efficiency and $t_{dyn} \approx
\sigma/(3.5\pi G\Pi)$ is the dynamical time scale.

On the other hand, the supernova remnants (SNRs) will quickly
radiate their thermal energy when they propagate in the dense
environment such as the central gas reservior for starbursts, and
enter the momentum driven phase \citep{MN04}. For a typical value,
the net momentum deposited by supernovae has approximately the same
order as that deposited by the radiation pressure \citep{MQT05}. The
whole momentum deposition rate can be written as \beq \dot P=\dot
P_{rp}+\dot P_{sn}=\xi_m\epsilon\dot M_{\star}c,\eeq where
$\xi_m=1+\dot P_{sn}/\dot P_{rp}$ is assumed of order of unity in
our model.

\subsection{The halo-regulated star formation}

Gas in the inner region has higher density and shorter dynamical
time, so the star formation feedback should be stronger there than
in the outer region. The total net star-forming feedback is
outward at large scale (however, we must bear in mind that it may
not be the case at very small scale, see discussions in \S4), and
the outward star-forming feedback can be regarded as the
anti-gravitational source. Because the gas density in the starburst
region is extremely high and the cooling of the gas is very
efficient, the heating effect is neglected. The star-forming
feedback could increase the potential energy of the
whole system so as to help the system back to the virial
equilibrium. We assume that the whole system re-virialize at radius
$r'$.  According to the above argument, we expect that the additional
inward drag provided by the background dark matter potential is
compensated by the outward star-forming feedback. Namely, we should
have \beq \int\xi_m\epsilon\dot M_{\star}cdr'=\xi_m\epsilon\dot
M_{\star}cr'=\pi\mathit{\Pi}GM_{c,max}r',\eeq where the left side
and right side of Eq. (10) denote the equivalent potential energy of
the star forming feedback and the potential energy from the dark
matter background. Here we use the fact that the quantity
$\xi_m\epsilon\dot M_{\star}c$ is independent on the radius $r'$ and
we assume that the stars form deeply inside the molecular cloud and
don't appear optically bright during the main epoch of star
formation. Eq. (10) can also be directly written as
\beq\xi_m\epsilon\dot M_{\star}c=\pi\mathit{\Pi}GM_{c,max}.\eeq
The above equation suggests that the star formation rate of the
central starburst region, being directly related to the total collapse
gas mass $M_{c,max}$, is constant if $M_{c,max}$ is fixed. If
the star formation rate increases slightly, the left side of Eq. (11)
will be larger than the right side, and the system is puffed up due to the star-forming feedback, and the star formation rate will therefore
decrease again to make the whole system return to the equilibrium
state. So Eq. (11) requires that the star formation activity in the
central starburst region is regulated by the gravity from the dark
matter background. Such "halo-regulated" star formation mode is very
different from that in the nearby disk galaxies. Here we calculate
some parameters based on this mode in order to compare them with the
observations.

Using Eq. (8) to eliminate $\dot M_{\star}/M_{c,max}$ in Eq. (11),
from the expression of $t_{dyn}$ in \S3.1 we obtain the star
formation efficiency as \beq \eta=\frac{\sigma\pi}{3.11\xi_m\epsilon
c}=0.68\sigma_{200}\xi_m^{-1}\epsilon_3^{-1},\eeq where
$\epsilon_3=\epsilon/10^{-3}$. Substituting Eq. (7) into Eq. (8) and
using Eq. (12), we obtain the threshold star formation rate in the
central starburst region as \beq \dot
M_{\star}=\frac{\sigma^4}{1.1G\xi_m\epsilon
c}\approx1000\sigma_{200}^4\xi_m^{-1}\epsilon_3^{-1}M_{\odot}yr^{-1},\eeq
and the star formation timescale \beqa
t_{\star}&=&\frac{t_{dyn}}{\eta}=\frac{\xi_m\epsilon
c}{\pi\Pi G}\nonumber\\
&\approx&1.0\times10^8\xi_m\epsilon_3M_{v,12}^{-0.07}\zeta_c^{-1}(z)
yr.\eeqa

It is interesting to note that the star formation efficiency (SFE)
as well as the star formation rate (SFR) are closely related to the
velocity dispersion. The large the velocity dispersion is, the
higher SFR and SFE are. The result implies that there is higher
fraction of gas converting into stars in more massive galaxies (eg.
giant elliptical) and the high redshift SMGs observations show that
it may be just the case \citep{AX05}.

\section{THE INWARD STAR FORMING FEEDBACK AND THE OBSCURED BLACK HOLE GROWTH}

\begin{figure}
\epsscale{0.5} \plotone{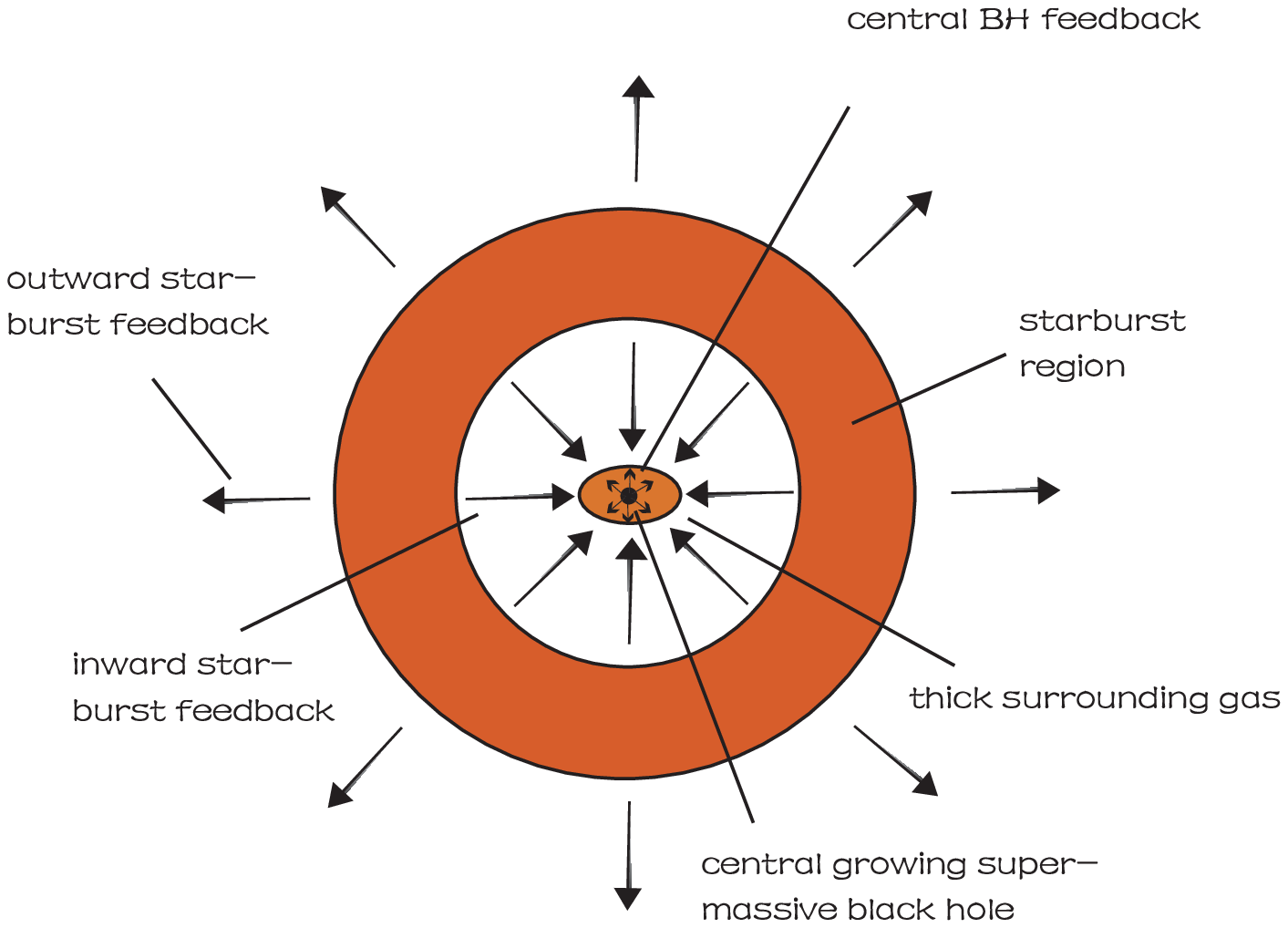}
 \caption{A cartoon which describes the obscured growth of the central BH at the early growth stage. The shaded part is starburst
 region, which generates momentum feedback in two directions if there are enough dusts. The outward one resists the gravity while
 the inward one regulates the growth of the BH. Because the mass of BH is small at early stage, the feedback from BH
 can't balance the inward feedback from the starburst. The push of the inward starburst feedback combining with the
remained angular momentum force the gas to form the above shape. The
BH will hide in the thick gas shell and can not be seen in optical
band.} \label{fig:bb}
\end{figure}

Because the star formation is unlikely to proceed efficiently at
very small scales (e.g. galactic nuclei), the starburst region can
not be regarded as a point source comparing with the
central BH, and the momentum feedback from starbursts should
transport in two directions: the outward one to resist the gravity
and the inward one to drive some part of gas to feed the BH. If the
initial BH mass is not very large ($\le10^6M_{\odot}$), the inflow
gas is sufficient for the BH to accrete at Eddington accretion rate
for a long time. Initially, the feedback exerting on the surrounding
gas of the BH can't balance the inward feedback from the starburst
region, the BH will hide in a gas shell (see Figure~\ref{fig:bb} for
a cartoon view) and be optically thick to the optical and UV
radiations. Hence the main growth phase of the BH should
be obscured (so-called "pre-quasar" phase), which is consistent with
the SMG observations \citep{AX05}.

The feedback from BH will be stronger as the BH grows bigger.
Because of the obscured growth environment and the possible high
accretion rate, the momentum flux may transport outwards via a
Compton-thick wind launched from the BH's accretion disk. If we
assume that the covering factor of the wind is large enough, the net
momentum flux rate deposited in the wind driven by the radiation
pressure from BH can be expressed as \citep{KP03} \beq
\dot{P}_{wind}=\dot{M}_{out}v_{out}=\frac{2L_{EDD}}{c}=\frac{8\pi
GM_{BH}}{\kappa}, \eeq where $\dot{M}_{out}$ is the outflow rate and
$v_{out}$ is the outflow velocity. Once the balance between the
inward starburst feedback and outward BH feedback is achieved, we
have \beq\dot{P}_{wind}=\xi_m\epsilon\dot M_{\star}c,\eeq then the
BH's feedback is large enough to halt the further gas supply, so we
can say that a BH will end its main growth phase after Eq. (16) is
satisfied. Based on the argument in \S2, the bulge mass is
approximated as $M_{c,max}$. Using Eqs. (11), (15) and (16), the
ratio of BH mass to bulge
 mass can be expressed as
\beq\frac{M_{BH}}{M_{bulge}}=\frac{\kappa\mathit{\Pi}}{8}\approx1.4\times10^{-3}
M_{v,12}^{0.07}\zeta_c(z).\eeq Combining Eq. (17) with (7), we can
also derive the $M_{BH}-\sigma$ relation \beq
M_{BH}\approx\frac{\kappa\sigma^4}{10\pi
G^2}=1.8\times10^8\sigma_{200}^4M_{\odot}.\eeq

The above results are surprisingly consistent with the local
observations \citep{MA98,MD02,MH03,TMA02}.

\section{DISCUSSIONS}

The so-called "halo-regulated" star formation mode requires the
additional gravity from dark matter to regulate the SFR and SFE to a
very high level (although they may drop at the late stage).  It
means that a large fraction of gas is converted into stars in a
relatively short timescale without being dispersed or blown out. The
SFR is much higher than that in normal disk galaxies inferred from
the Kennicutt Law, but is consistent with the observations of
high-redshift starburst galaxies \citep{SV05} and some small scale
star formation phenomena (e.g. SFE during the formation of the bound
cluster \citep{LMD85}). This convinces us that such star formation
mode actually works in extreme environments, such as the central
starburst region at high redshift, which is very different from that
of local universe. We also note that the star formation timescale
only has weak dependence with the redshift (see Eq. (14)), and is
independent on the mass of the galaxy roughly keeping the same over
a large redshift range. It may explain the narrow range
($100\sim300Myr$) of the starburst duration inferred from
Faber-Jackson relation \citep{MQT05}.

 Many previous momentum driven models argue that when the BH grows
 big enough it can blow most of the gas away to
 end its growth \citep{KI03,MQT05,BN05}.
 The main difference between our model and these previous models is that in our model the inward star forming feedback plays important role in
 regulating the BH mass. Simulation shows that the
 inward star forming feedback transports most momentum but little gas into
 the nuclear region \citep{MFM02}. Hence, the gas obscuring
 the BH is much less than the total amount of gas in the starburst region.
 According to Eq. (11), we find that the gravity of the gas in the
 nuclear region is much smaller than the inward star forming feedback. So
 it is reasonable to reckon that the BH is confined by the inward star forming feedback rather than to drive
 large scale outflow. The BH will end its growth once the feedback from BH is big enough to
 balance the inward star forming feedback, and the BH-bulge relation is
 naturally derived from the final balance condition (see Eq.(16)).
 There are two interesting points to note: first, our derived $M_{BH}-\sigma$
 relation (see Eq.(18))
is independent of the parameter of the detailed star formation
 process; second, the derived mass ratio between the BH and bulge is only related to some
 properties of the background dark matter density profile $\mathit{\Pi}$, independent
 of the progenitor's gas fraction or the total gas mass
 in the nuclear region. We think that it is all because that the "halo-regulated" star formation mode makes the
 star formation activity as an equivalent effect as the dark matter gravitational effect.
 We also find that our derived $M_{BH}-\sigma$ relation (see Eq.(17)) has no evolution with the redshift, which is
 consistent with many previous models and the observations out to the median redshift \citep{SH03,WU07}, while the
 $M_{BH}-M_{bulge}$ relation has weak
 dependence on the redshift. At high redshift $z\sim3$,
 $M_{v,12}^{0.07}\zeta_c(z)$ change by
 a factor $10^{0.14} \times 0.5^{2/3} \times 1.4 \sim 1.3$;
 assuming $M_{vir}$ was ten times smaller at high redshift.  From Eq. (17)
 we can see that the total effect is: the $M_{BH}/M_{bulge}$
 ratio is roughly the same, only with small scatters (at $z\le3$).
 The result implies that at least some part of the scatter of the $M_{BH}-M_{bulge}$ is
 intrinsic.

\begin{figure}
\epsscale{0.5} \plotone{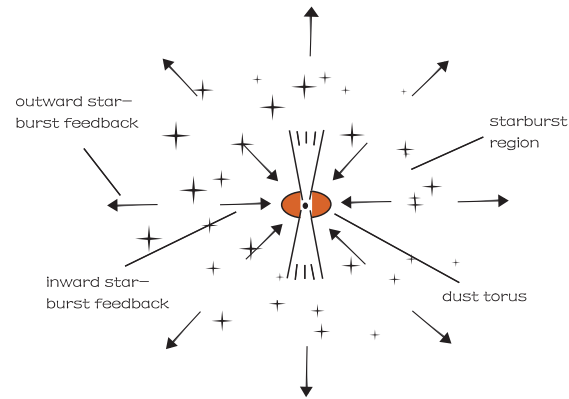}
 \caption{A cartoon which describes the stage after the main (obscured) growth stage of the central BH. Till the feedback from the BH balances
 the inward feedback from the starburst, compton-thick wind from BH can easily "crush the cocoon" in the direction of the angular momentum and form
 the torus-shape. It may explain the formation of the dust-torus in many AGNs.}
\label{fig:ab}
\end{figure}

Furthermore, the remaining angular momentum of the dusty gas
surrounding the BH during its obscured growth phase may not be
negligible at very small scale. This angular momentum could make the
configuration of the shrouded gas deviate from the ideal sphere, and
distribute more like a "hamburger". Therefore, once the BH grows
large enough, the wind from the direction of angular momentum will
first "crush the cocoon" and show an optically bright phase (see
Figure~\ref{fig:ab} for a cartoon view). Such a scenario offers a
natural explanation to the formation of the dust-torus in AGNs,
which is considered as an important part in the AGN unification
model \citep{AN93}. We also predict that the dust in the dust-torus
of AGN may originate from the outside starburst region, and the
covering angle of the dust torus may closely relate to the BH mass.
This tentative result needs to be confirmed by the future
observations.

\section{CONCLUSIONS}
Here we itemize the conclusions which we have derived in this
paper:

1. We suggest that bulges form in a "monolithic" collapse and
subsequent rapid star formation. We argue that such runaway collapse
is due to the gravothermal instability. The isothermal sphere with a
NFW dark matter background becomes unstable when the central gas
pressure is about 30 times larger than the central dark matter
pressure.  The collapsed gas will result in vigorous star formation
activities and is the direct material source of the bulge formation.

2. The squeeze and compression help the gas to form molecular
clouds, then many stars form in those clouds during a relatively
short timescale and the central starburst is triggered. Intense
starburst brings strong star forming feedback, having the
$M_{g,min}$ and $r_c$ in hand and assuming the momentum driven
mechanism dominates, we calculate the star formation rate and the
related star forming feedback. We argue that the star forming
feedback serves as the source to resist the gravity and will finally
help the system return to equilibrium and halt the gas collapse.
Hence, the "halo-regulated" star formation mode is required and we
find such star formation mode can well match the high redshift
starburst observations.

3.  We reckon that the inward star forming feedback should exist to
conserve the local momentum and to connect the outside starburst
region and nuclear region. The inward star forming feedback can push
large amounts of gas to the central region to obscure and confine
the central BH (if the BH mass is not so large) and the BH system
may appear as a "pre-quasar". We suggest that the BH growth is just
regulated by the inward star forming feedback. We give the condition
for the BH to end its growth is that the compton-thick disk wind
balances the inward star forming feedback. Based on this we derive
the BH-bulge relations which are well consistent with the local
observations.

\acknowledgments
The authors thank Fukun Liu and Shude Mao for
helpful discussions and the anonymous referee for helpful
suggestions. XBW acknowledges the support of NSFC Grant (No.10473001
and No.10525313) and the RFDP Grant (No.20050001026).

\clearpage

\end{document}